\documentclass{iau}
\usepackage{graphicx}

\title[S 295.~~Future Observational Prospects] %% give here short title %%
{Future prospects in observational galaxy evolution: towards increased resolution.}

\author[K. Glazebrook]   %% give here short author list %%
{Karl Glazebrook$^1$
 }

\affiliation{$^1$Swinburne University of Technology, PO Box 218, Hawthorn, Vic 3122, Australia \\ email: {\tt kglazebrook@swin.edu.au} 
}

\pubyear{2013}
\volume{295}  %% insert here IAU Symposium No.
\pagerange{xx--xx}
\jname{The intriguing life of massive galaxies}
\editors{D. Thomas, A. Pasquali \& I. Ferreras, eds.}
\begin{document}

\maketitle

\begin{abstract}
Future prospects in observational galaxy evolution are reviewed from a personal perspective. New insights will especially come from high-redshift integral field kinematic data and similar low-redshift observations in very large and definitive surveys. We will start to systematically probe the mass structures of galaxies and their haloes via lensing from new imaging surveys and upcoming near-IR spectroscopic surveys will finally obtain large numbers of rest frame optical spectra at high-redshift routinely. ALMA will be an important new ingredient, spatially resolving the molecular gas fuelling the high star-formation rates seen in the early Universe.
\keywords{galaxies: evolution, galaxies: formation, galaxies: high-redshift, telescopes, instrumentation: miscellaneous }
\end{abstract}

\firstsection % if your document starts with a section,
\section{Introduction}

I would like to thank the organisers for their kind invitation to review the future observational prospects in galaxy evolution, and in particular for massive galaxies, the theme of this Symposium. I am going to attempt to look forward about five years, this seems a sensible time frame on which to make predictions of what will be the most highly impactful observations.

If we review \emph{the last five years }for comparison, it is quite startling to see the unexpected discoveries and developments that came about. Here are the ones that stick most in my mind (and references are intended to be illustrative not complete!):
\bigskip
\begin{enumerate}
\item The dramatic size evolution found in elliptical galaxies --- up to a factor of five since $z\sim 2$ (van Dokkum et al. 2008, Cimatti et al. 2008, Damjanov et al 2009).
\item The existence of an evolving star-formation rate--stellar mass `main sequence' for star-forming galaxies (Noeske et al. 2006).
\item That most stellar mass growth in massive galaxies occurs via \textit{in situ} star-formation and not via mass delivery in mergers (Conselice  et al. 2012).
\item That massive star-forming galaxies at $z\sim 2$ show a large fraction of rotating disks (Genzel et al. 2006).
\item That the clumpy morphologies of high-redshift galaxies are likely due to giant star-formation complexes driven by the Jean's scale in turbulent high-velocity dispersion disks (Bournaud et al. 2009).
\item That the universality of the Initial Mass Function (IMF) is now back in question (van Dokkum 2010, Hoversten \& Glazebrook 2008).
\item That the various physical properties of galaxies on the `red sequence' or `blue cloud' seem to be set solely by their stellar mass and to be independent of environment (e.g. Balogh et al. 2004,  Baldry et al. 2006, Moucine, Baldry \& Bamford 2007, Mocz et al. 2012, Peng et al 2010, Thomas et al. 2010), i.e. the only effect of environment seems to be in setting the numbers of red vs blue objects, perhaps via a threshold effect.

\end{enumerate}

\bigskip
Given the recent history of unexpected developments in galaxy evolution this seems to make predicting the next five years fairly perilous! One thing that makes it slightly easier is that no major new telescopes will be commissioned during the period, indeed the new generation of Extremely Large Telescopes (ELTs) won't arrive until at least 2018. Other new large facilities such as the Large Synoptic Survey Telescope and the Square Kilometre Array are destined for the 2020's. 

In this look forward I am going to focus on three major areas that I have picked on due to upcoming new capabilities: (i) galaxy structures and kinematics, (iii) high-redshift imaging and spectroscopic surveys and (iii) the imminent revolution in sub-mm astronomy from the Atacama Large Millmetre Array (ALMA). 

\section{Galaxy Structures and Kinematics}

Integral Field Spectroscopy (IFS) has revolutionised the study of the kinematics of high-redshift star-forming galaxies and we now have about 100--200 high-quality observations of galaxies at $z\gtrsim 1$ from various surveys and nicely reviewed in S. Wuyt's talk at this Symposium. At these redshifts we see a picture where galaxy kinematic classes appear three-way split into (i) rotating objects with clearly disk-like velocity fields (ii) objects with kinematic structures but no uniform disk-like pattern (sometimes said to be `mergers') and (iii) objects with no kinematic structure (sometimes referred to as `dispersion dominated', Law et al. 2007). The split here is around 20--40\% in each class but this is sensitive to the particular survey and selection function and the fraction of disks seems to increase towards higher stellar masses (F\o rster-Schreiber et al. 2009). Objects with disk kinematics seem to follow a Tully-Fisher relation in that they have the tightest scatter around a luminosity (or stellar mass) vs circular  velocity line with a similar slope to, but a small offset from, the local Tully-Fisher relation (Puech et al. 2008, Cresci et al. 2008). A particular development at this symposium is a nice Tully-Fisher relation at $z\sim 1.2$ form the MASSIV survey (P. Amran talk), a redshift in which there was previously somewhat of a gap.

One key upcoming development is the advent of the KMOS IFS (Sharples et al. 2004) which is to be commissioned on the Very Large Telescope  at the end of 2012. This offers the first near-IR multiplexed IFS on a large telescope and IFS observations of up to 24 galaxies can be performed simultaneously. This will enable two important advances: first, and obviously, much larger high-redshift IFS kinematic samples will be obtainable allowing statistical trends to be studied. Secondly the large multiplex means it will be efficient to study much fainter galaxies with longer exposure times. Current IFS surveys are restricted to studying the more luminous (in emission lines) objects, typically around $\sim L^*$ in H$\alpha$ at $z\sim 2$, thus being able to tackle even small numbers of sub-$\L^*$ objects will allow selection biases to be studied. 

One open question, in my mind, to be tackled by future surveys is the evolution of the galaxy merger rate. IFS surveys typically identify 20--30\% of galaxies as mergers via kinematics at $0.5<z<2$ (Yang et al. 2008, Lopez-Sanjuan et al. 2012, F\o rster-Schreiber et al. 2009) which is in stark contrast to the local value of $\sim 4\%$. Are the merger rates identified via kinematics consistent with those measured by close-pair counts (e..g Y. Peng, this Symposium)? Can we even objectively identify mergers in kinematic maps? Pioneering work in this latter topic was done using kinemetry by Shapiro et al. (2008) but needs to be further developed, especially with respect to local calibration samples. In this Symposium P. Amran showed a new and different approach to quantitively identifying mergers. This is an excellent area for the future development of parametric and non-parametric statistics. A related question is can we go to the next step and measure mass ratios and merger timescales from IFS maps?

\begin{figure}[t]
% \vspace*{-2.0 cm}
\begin{center}
 \includegraphics[width=3.4in]{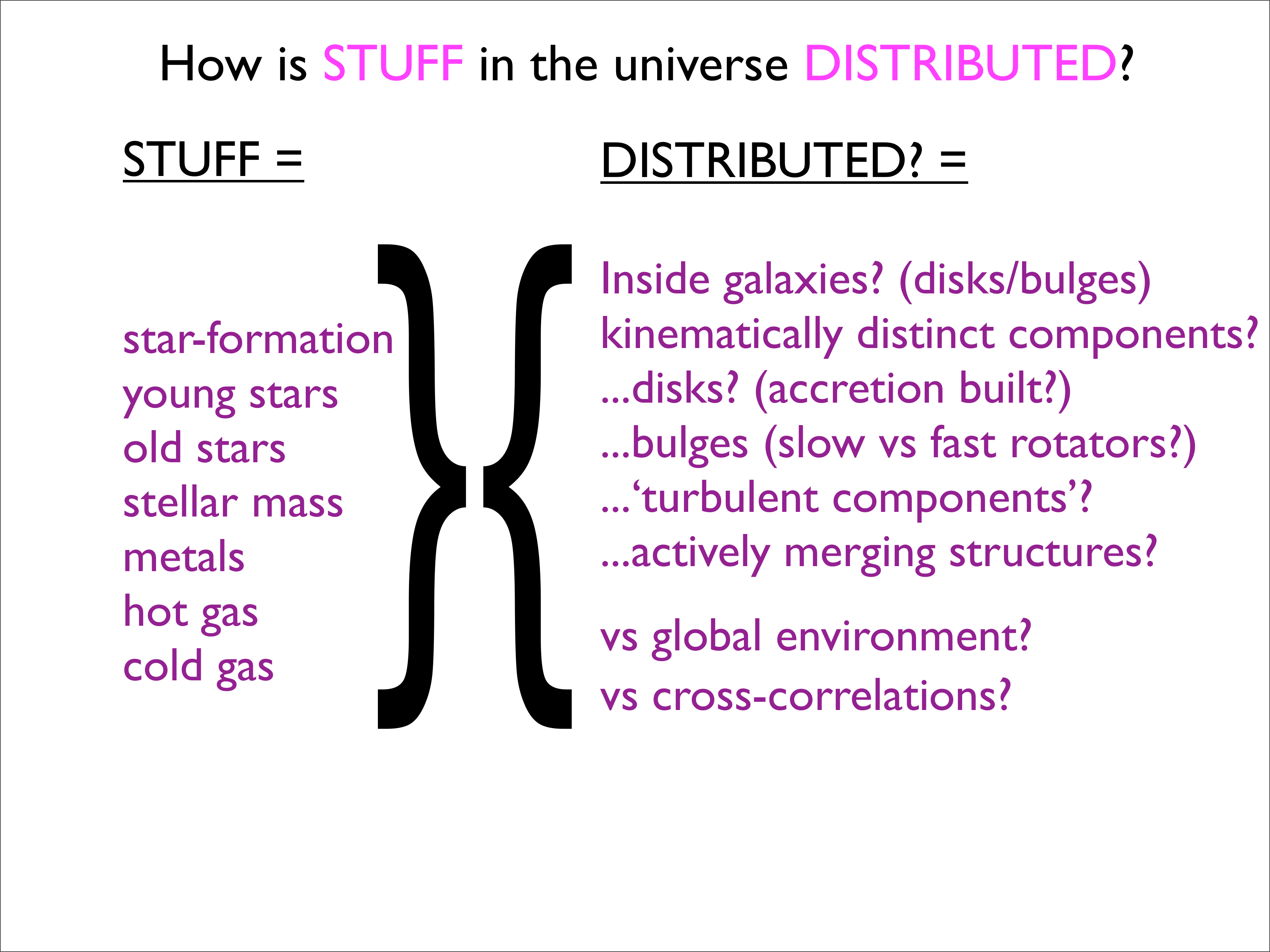} 
% \vspace*{-1.0 cm}
 \caption{The meta-questions of IFS surveys. What is the mapping between the left and the right columns?}
   \label{fig1}
\end{center}
\end{figure}

I believe the other key development will be the carrying out of large-scale local IFS surveys, a `kinematic SDSS'. Current local IFS samples are of order several hundred galaxies, diversely selected and with heterogeneous data. This is analogous to the situation for imaging and 1D spectroscopic surveys before the 2dF Galaxy Redshift Survey and the Sloan Digital Sky Survey (SDSS). The next five years will see surveys of several thousand, perhaps tens of thousands of local galaxies done with multiplexed IFS instruments. Projects actively building instruments and planning observational campaigns in the near term are the SAMI consortium (Croom et al. 2012), who will use the Anglo-Australian Telescope, and the MANGA team (P.I. Kevin Bundy) planning to use the SDSS telescope. These instruments typically deploy $\sim$ 20 integral field units in a 2--3$^\circ$ field-of-view. This will allow the statistical study of the distribution of resolved kinematic structures in the local Universe and other meta-questions (Figure~1). In particular we will move away from scaling relations such as Tully-Fisher to the study of true kinematic distribution functions where space-density plays a key role in comparing with theoretical models. These surveys will also provide a cornerstone for quantitative comparison with high-redshift surveys, for example by providing a high-quality merger sample where mergers are identified by kinematics and photometry (e.g. tidal tails and other low surface brightness features that may not be visible at high-redshift). They can also be used to find rare local analogues of high-redshift galaxies: because they are nearby they can then be followed up in exquisite detail to see what makes the tick astrophysically. One example of this is the work of Green et. al 2010 where we identified candidate local turbulent disks with high star-formation rates. We are currently engaged with HST, Gemini IFS and other facilities to prove if they are indeed analogues and how the star-formation is driven.

We have also seen some nice work presented in this symposium on the kinematics and structures of red galaxies from high to low redshift. The so-called `two-phase model' for the assembly of red galaxies (Forbes et al. 2011, Figure~2) is becoming popular where red galaxies start out as compact and very dense primordial `red nuggets'\footnote{Confession: my own invented phrase, now seems increasingly apt!} and then accrete a stellar halo via minor mergers as the core loses density. This allows a considerable amount of evolution of effective size per unit stellar mass increase and seems to be the emerging consensus explanation of size evolution in red galaxies. This does beg the question as to how the initial red nugget forms, is it via dissipative monolithic collapse and rapid starburst of a primordial gas cloud? Or the quenching or merging of high-redshift disks? Is this consistent with the axial ratios and Sersic indices being found at high redshift? (e.g. Damjanov, this symposium, Chevance et al. 2012.) We now have a limited number of velocity dispersion measurements, from absorption lines, of the most massive high-redshift ellipticals which seem to supper the minor-merger hypothesis (e.g. I. Trujillo's review in these proceedings). What we do not yet have is \emph{resolved} kinematic measurements, for example are the red nuggets very rapidly rotating disks?  Absorption line measurements are very difficult but future deep IFS observations such as those of KMOS can address this question. So will deep imaging using multi-conjugate adaptive optics (AO) which will deliver resolution 2--3$\times$ that of HST (McGregor et al. 2004). 

At low redshift it remains to be seen if the two-phase model can reproduce the distribution of elliptical galaxies between slow and fast rotators which has now been measured in the field and in very dense environments (R. Davies, these proceedings). Does the real cosmological merger history deliver the right final angular momentum distribution? This is a challenge for theory as well as observers (e.g. Burkert et al. 2008). Surveys such as MANGA and SAMI will deliver much better statistics but hydrodynamic simulations of massive galaxies embedded in large cosmological volumes remains supercomputer-intensive.

\begin{figure}[b]
% \vspace*{-2.0 cm}
\begin{center}
 \includegraphics[width=3.4in]{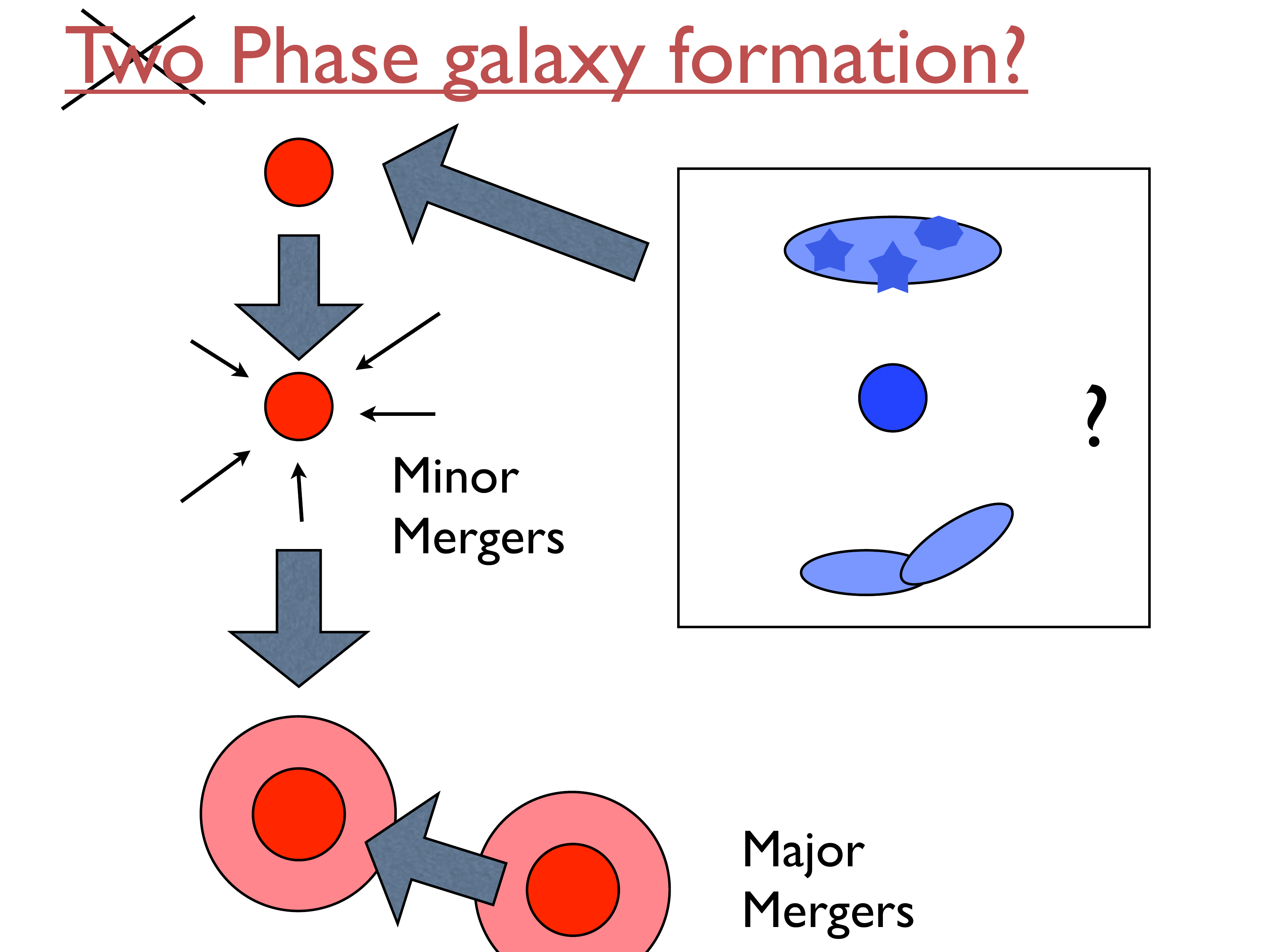} 
% \vspace*{-1.0 cm}
 \caption{The `Two-Phase Model' of galaxy formation? A red nugget at $z\sim 2$ grows a stellar halo and a considerable size increase via minor mergers. In some cases it may undergo major mergers to build a massive red galaxy. But what is Phase Zero? How does the red nugget get there in the first place from some blue predecessor? Possible mechanisms include fading of a clumpy disk, of a blue nugget or from a disk merger are illustrated. All would predict different spatial and kinematic morphologies for the red nugget.}
   \label{fig2}
\end{center}
\end{figure}

One final question that is perhaps unlikely to be answered in the next five years is the nature of the dispersion dominated compact \emph{star-forming} galaxies that seem to constitute almost a third of the population. These are lower mass ($<5\times 10^{10} M_\odot$) so may not be related to the red nuggets even though they are a similar size ($\sim $1--2 kpc). Are they purely dispersion dominated or do these conceal very compact disks that are unresolved even with AO IFS? This may require AO on ELTs to resolve, though spectroastrometry (Gnerucci et al. 2011) may allow information to be gleaned in the nearer term.

\section{High-Redshift Imaging and Spectroscopic Surveys}

In the last five years it has become routine for deep optical imaging surveys ($AB\sim 26$--27) to cover tens to hundreds of square degrees. At these depths galaxies are surveyed to $z\sim 6$. In the next five years even more gigapixels on sky will allow surveys such as the Dark Energy Survey (Flaugher 2005) and the Hype Suprime-Cam survey (Takada 2010) to cover thousands of square degrees at these depths. VISTA will similarly allow deep and wide near-IR surveys (McPherson at al. 2004). As outlined by D. Capozzi in these proceedings these imaging surveys will contribute to galaxy evolution studies via accurate measurements of photometric redshifts, luminosity functions, galaxy clustering, etc.

However at the risk of some controversy I predict that the most important applications to galaxy evolution from the new imaging surveys will come from the use of galaxy lensing enabled by such large areas. Weak lensing will enable the direct statistical measure of dark matter in galaxy and cluster haloes --- some very nice work along these lines using the CFHT Legacy Survey was presented by M. Hudson in these proceedings showing a good correlation between dark halo mass and stellar mass fraction in red and blue galaxies very suggestive of possible physical mechanisms. 
Strong lensing is also very powerful especially when combined with kinematic data (e..g. T. Treu talk in this symposium) as it allows mass structures and the IMF to be measured in the lensing galaxy. It is also very good for studying the lensed galaxy due to the large magnification of the light, making it brighter but also allowing smaller spatial scales to be resolved if the lens model can be inverted. The prospects of wider imaging surveys contributes to both weak lensing, via better statistics, and to strong lensing allowing more of these rare phenomena to be found.

In spectroscopy the instrument that I am personally most excited about is MOSFIRE, the near-IR multislit spectrograph commissioned on Keck in mid-2012. This cryogenic instrument operates from 0.9--2.4$ \mu$m and allows slit spectroscopy of up to 46 targets simultaneously (McLean et al. 2011). In my view it offers the first combination of three key features required to make near-IR spectroscopy succeed for faint high-redshift targets: (i) sufficient 
spectral resolution ($R=3300$) to well-resolve the airglow OH background out and `get between the lines'. (ii) low scattered light and thermal background meaning it is truly dark between the sky lines; the measured interline background of MOSFIRE is very dark and comparable to the measurements of Maihaira  et al.(1993). (iii) low readout noise and (iv) high instrument throughput 30--40\%. Other similar instruments exist (such as F2 on Gemini) but do not offer the same spectral resolution for the one arcsec slit sizes required and have yet to be demonstrated on sky. The performance of MOSFIRE is shown by the detection of H$\alpha$ in normal Lyman Break Galaxies at $z\sim 2$ in exposure times as short as 30 minutes! \footnote{See `first light presentation' on http://irlab.astro.ucla.edu/mosfire/}

The key science area which will be tackled by MOSFIRE is the routine continuum spectroscopy of normal galaxies at high-redshift in large numbers in the rest-frame optical for detailed comparison with low redshift surveys such as SDSS. These spectra will measure spectroscopic redshifts, stellar populations, metallicities and velocity dispersions for homogenous samples.  Without an instrument such as MOSFIRE this has been very difficult and most work in the last decade has relied on photometric redshifts.  Even the very simplest product --- redshift --- should not be ignored as it allows clusters, environments and larger scale structures to be defined at high-redshift. These are the context of high-redshift galaxy evolution and current spectroscopic samples are highly biassed towards subsets of the population such as Lyman Break Galaxies. Photometric redshifts do not have the accuracy to measure such 3D environments though the most accurate ones, with medium band filters, do start to identify large scale structures and clusters (Spitler et al. 2012, Labb\'e\ talk this symposium) but require spectroscopy to confirm. The prospects for MOSFIRE surveys are excellent with high-quality very deep high-quality near-IR imaging data for selection already available from HST (the CANDELS survey, Grogin et al. 2011, Koekemoer et al. 2011) and from the ground with medium bands. Because of this nexus we will now see a renaissance in high-redshift spectroscopy. It is interesting to note that this capability was in fact a key  original science goal of 8m class telescopes and in the next five years we will finally see it delivered.

Towards the end of the five year forecast we may see the Subaru Prime Focus Spectrograph arrive (Ellis et al. 2012) offering a 50-fold increase in optical near-IR multiplex and field-of-view over current systems (though being non-cryogenic will operate at wavelengths $<1.5 \,\mu$m). This will open the exciting prospect of using galaxies at $z>>1$ for {\em cosmology} as well as galaxy evolution.

\section{The Age of ALMA}

As I write one very significant new telescope is being commissioned: ALMA (Hills \& Beasley 2008). Virtually no ALMA results were presented at this symposium as very few people actually have any ALMA data.\footnote{A show of hands at the symposium revealed at most 2--3 hands up in the audience.} So far no more than about 1000 hours of ALMA science time has been available to the community. However if we have a conference such as this in five years time I fully expect ALMA results to dominate the conference.

Why do I say this? Today high-redshift is dominated by optical and near-IR observations which are mainly sensitive to stars and hot ionised gas (e.g. from star-formation or AGN). However we need to consider the fuel as well as the fire. We know from current sub-mm observations that the molecular gas fractions of massive galaxies rises from a mere 5--10\% at $z=0$ to $\sim 50\%$ at $z\sim 2$ (Daddi et al. 2010,  Tacconi et al. 2010). This probably accounts for the high prevalence of unstable, clumpy,  turbulent disks (e.g. Genzel et al. 2008) and necessitates high inflow rates of cosmic material to sustain them (Dekel et al. 2009). 

However current sub-mm telescopes barely resolve high-redshift galaxies with 0.5--1 arcsec beams and require many hours of integration per target. ALMA will improve this by factors of ten and enable kpc-resolution morphology and kinematics of molecular gas and dust in normal star-forming galaxies to be routinely made. We predict the clumpy disks to be gas rich and thick. Will we see thick cold molecular gas disks co-rotating and aligned with the young stars seen by the near-IR IFS observations? Will we see {\it super-giant molecular clouds} associated with the giant star-forming regions see in the UV? I predict we will!

A particularly important question for ALMA's spatial resolution is the nature of the star-formation law relating gas density to star-formation rate, a critical theoretical ingredient of galaxy formation simulations (the `sub-grid physics').  Around 80\% of the stars in the Universe formed at $z>1$ but we have seen throughout this conference that  galaxies in the the high-redshift Universe are very different to today. Will the star-formation law be the same or quite different? The classical Kennicutt-Schmidt law (Kennicutt 1998) simply relates surface densities of gas and star-formation via a power law. Even locally there are many variations on this theme (a topic extensively discussed in Symposium 292 the previous week), for example there may be `thresholds' or a volumetric relation may be more appropriate (Krumholz, McKee \& Tomlinson 2009).  At high-redshift Daddi et al. (2010) suggested there are  in fact two relations --- a `sequence of starbursts' and a `sequence of disks' but which may be unified by introducing a dynamical time in to the formulation. ALMA will bring a highly superior set of data to bear on this problem and I will predict some surprises!

Finally one interesting prediction that could perhaps be tested by ALMA is the existence of {\em dark} turbulent disks (Elmgereen \& Burkert 2010). The prediction is that turbulence in gas disks starts initially in an accretion driven phase lasting for $\sim 180$ Myr before star-formation turns on. The gas would be cold and molecular --- the visibility of such objects to ALMA has not yet been calculated, but would make for an interesting paper.

\section{Final Words}

Some firm predictions for the next five years:

\begin{enumerate}
\item We will see a move back to real spectroscopic surveys at $2<z<5$.
\item A `Golden Age' of Integral Field Spectroscopy of large samples including definitive local surveys.
\item We will probe the `fuel for the fire' with ALMA.
\item We will {\em still\/} be arguing about stellar population synthesis model ingredients (if this conference is anything to go by!).
\end{enumerate}

Finally it is amusing to note that at this conference we saw Carlos Frenk (doyen of semi-analytic modelers) saying that `galaxy formation is complicated' and Simon Lilly (the archetypal observer) saying `galaxy formation is simple'! This appears to be a reversal of the theory-observer dichotomy of ten years ago to my memory, however I will dare to suggest that they are both in fact wrong! I think in the next 5--10 years we will see basic physical questions of star-formation and quenching (i.e. the formation of the red sequence) ironed out through better spatially-resolved observations as described above and there will be less need for `recipes' in both camps. I speculate these observations will reveal new simplicities but also more complexity then the over-simplified picture that has arisen from large surveys with integrated spectra.


\begin{thebibliography}{}

\bibitem[]{} Baldry, I. K., Balogh, M. L., Bower, R. G., Glazebrook, K., Nichol, R. C., Bamford, S. P., Budavari, T., 2006,  \textit{MNRAS}, 373, 469

\bibitem[]{} Balogh, M. L., Baldry, I. K., Nichol, R., Miller, C., Bower, R., Glazebrook, K., 2004, \textit{ApJ}, 615, L101-L104

\bibitem[]{} Bournaud F., Elmegreen B. G., 2009,  \textit{ApJ}, 694, L158 

\bibitem[Burkert et al.(2008)]{2008ApJ...685..897B} Burkert, A., Naab, T., 
Johansson, P.~H., \& Jesseit, R.\ 2008, ApJ, 685, 897 

\bibitem[]{} Chevance M., Weijmans A.-M., Damjanov I., Abraham R. G., Simard L., van den Bergh S., Caris E., Glazebrook K., 2012,  \textit{ApJ}, 754, L24

\bibitem[Cimatti et 
al.(2008)]{2008A&A...482...21C} Cimatti, A., Cassata, P., Pozzetti, L., et al.\ 2008, A\&A, 482, 21 


\bibitem[]{} Cresci et al. 2009,  \textit{ApJ} (2009) vol. 697 pp. 115

\bibitem[]{}  Conselice, C.~J., 
Mortlock, A., Bluck, A.~F.~L., \& Gruetzbauch, R.\ 2012, \textit{MNRAS}, in press, arXiv:1206.6995 

\bibitem[]{}   Croom, S.~M., Lawrence, 
J.~S., Bland-Hawthorn, J., et al.\ 2012, \textit{MNRAS}, 421, 872 

\bibitem[]{} Daddi, E., Elbaz, D., 
Walter, F., et al.\ 2010, \textit{ApJ}, 714, L118 


\bibitem[]{} Damjanov, I., et al. 2009, \textit{ApJ}, 695, 101

\bibitem[]{} Dekel A., et al., 2009, \textit{Nature}, 457, 451

\bibitem[Ellis et al.(2012)]{2012arXiv1206.0737E} Ellis, R., Takada, M., 
Aihara, H., et al.\ 2012, \textit{Extragalactic Science and Cosmology with the Subaru Prime Focus Spectrograph (PFS)}, arXiv:1206.0737 

\bibitem[]{}  Flaugher, B.\ 2005, 
\textit{International Journal of Modern Physics A}, 20, 3121 


\bibitem[]{} Forbes, D.~A., Spitler, 
L.~R., Strader, J., et al.\ 2011, \textit{MNRAS}, 413, 2943 

\bibitem[]{} F\o rster Schreiber N. M., et al., 2009,  \textit{ApJ}, 706, 1364


\bibitem[]{} Genzel R., et al., 2006, \textit{Nature}, 442, 786

\bibitem[]{} Gnerucci et al., 2011, \textit{A\&A},  533, 124

\bibitem[]{} Green A. W., Glazebrook K., McGregor P. J., Abraham R. G., Poole G. B., Damjanov I., McCarthy P. J., Colless M., Sharp R. G., 2010, \textit{Nature}, 467, 684

\bibitem[]{} Grogin, N.~A., Kocevski, 
D.~D., Faber, S.~M., et al.\ 2011, \textit{ApJS}, 197, 35 


\bibitem[]{}  Hills, R.~E., \& Beasley, A.~J.\ 2008, \textit{SPIE}, 7012

\bibitem[]{} Hoversten, E. A., Glazebrook, K., 2008, \textit{ApJ}, 675, 163
\bibitem[]{} Kennicutt R. C., Jr., 1998,  \textit{ApJ}, 498, 541 


\bibitem[]{}   Koekemoer, A. M. et al. 2011, \textit{ApJS}, 197, 36


\bibitem[]{} Krumholz M. R., McKee C. F., Tumlinson J., 2009, \textit{ApJ}, 699, 850 

\bibitem[]{} Law et al. 2007, \textit{ApJ}, 669, 929

\bibitem[]{} L\'opez-Sanjuan et al. 2012, \textit{AJ} in press (2012) arXiv:1208.5020

\bibitem[]{} Maihara et al., 1993, PASP, 105,  940

\bibitem[]{} McGregor P., Hart J., Stevanovic D., Bloxham G., Jones D., Van Harmelen J., Griesbach J., Dawson M., Young P., Jarnyk M. A., 2004, SPIE, 5492, 1033

\bibitem[]{} McLean I. S., et al.,  2010, SPIE, 7735

\bibitem[]{}  McPherson, A.~M., 
Born, A.~J., Sutherland, W.~J., 
\& Emerson, J.~P.\ 2004, \textit{SPIE}, 5489, 638 


\bibitem[]{} Mocz, P., Green, A., 
Malacari, M., \& Glazebrook, K.\ 2012, \textit{MNRAS}, 425, 296 

\bibitem[]{} Mouhcine M., Baldry I. K., \& Bamford S. P., 2007, \textit{MNRAS}, 382, 801

\bibitem[]{} Noeske, K. G., et al. 2007, \textit{ApJL}, 660, L43

\bibitem[]{}  Peng, Y.-j., Lilly, S.~J., 
Kova{\v c}, K., et al.\ 2010, \textit{ApJ}, 721, 193 

\bibitem[]{} Puech et al. , 2008,  \textit{A\&A}, 484,  173


\bibitem[]{} Shapiro et al., 2008, \textit{ApJ}, 682, 231

\bibitem[]{}  Sharples, R.~M., 
Bender, R., Lehnert, M.~D., et al.\ 2004, \textit{SPIE}, 5492, 1179 

\bibitem[]{} Spitler L. R., et al.,  2011, \textit{ApJ}, 748, L21

\bibitem[]{} Tacconi L. J.,et al., 2010,  \textit{Nature}, 463, 781

\bibitem[]{} Takada, M.\ 2010, \textit{American 
Institute of Physics Conference Series}, 1279, 120 

\bibitem[Thomas et al.(2010)]{2010MNRAS.404.1775T} Thomas, D., Maraston, 
C., Schawinski, K., Sarzi, M., \& Silk, J.\ 2010, MNRAS, 404, 1775 

\bibitem[]{} van Dokkum P. G., Franx M., Kriek M., Holden B., Illingworth G. D., Magee D., Bouwens R., Marchesini D., Quadri R., Rudnick G., Taylor E. N., Toft S., 2008, \textit{ApJ},  677, L5


\bibitem[]{} van Dokkum, P.~G., \& Conroy, C.\ 2010, \textit{Nature}, 468, 940 


\bibitem[]{} Yang et al., 2008,  \textit{A\&A},  477,  789


\end{thebibliography}
\end{document}